\documentclass[preprint,12pt]{elsarticle}

\usepackage{amssymb}
\usepackage{amsmath}
\usepackage{graphicx}
\usepackage{dcolumn}
\usepackage{bm}
\usepackage[mathlines]{lineno}
\usepackage{array}
\usepackage{multirow}
\usepackage{float}
\usepackage{longtable}
\usepackage{booktabs}
\usepackage[colorlinks = true,linkcolor=blue,citecolor=blue,urlcolor=blue]{hyperref}
\journal{surfaces and interfaces}

\begin{document}

\begin{frontmatter}



\title{Regulation of electronic structures in ReSeS monolayer with anisotropic deformations}


\author[A]{Timsy Tinche Lin}
\affiliation[A]{organization={National Laboratory of Solid State Microstructures, School of Physics, Nanjing University},
            city={Nanjing},
            postcode={210093}, 
            state={Jiangsu},
            country={China}}
\author[B]{Haochen Deng}
\affiliation[B]{organization={Key Laboratory of Micro and Nano Photonic Structures (MOE), Department of Optical Science and Engineering, Fudan University},
            city={Shanghai},
            postcode={200433}, 
            country={China}}
\author[A]{Junwei Ma}

\author[A]{Lizhe Liu}

\begin{abstract}
Because of their unique and rich physical properties, transition metal dichalcogenides  (TMDs) materials have attracted much interest. Many studies suggest that introducing the degree of freedom of anisotropy—which may be brought about by low structural symmetry—might further optimize their applications in industry and manufacturing. However, most currently reported TMDs do not achieve the theoretical minimum symmetry. Utilizing the first principles calculation, we present ReSeS monolayer with a Janus structure. Results indicate that its electronic dispersion is sensitive to structural distortions, which increases metallicity. Our reduction-Hamiltonian can provide a qualitative description, but further analyses reveal that bonding/antibonding properties near the Fermi surface are the more fundamental cause of the variations. Furthermore, geometric deformations can regulate the effective mass of electrons as well as the spectroscopic response, resulting in anisotropic behaviors. Our ideas serve as a foundation for developing new regulable optoelectronic devices.
\end{abstract}

\begin{graphicalabstract}
\begin{figure}
\centering
\includegraphics[width=14cm]{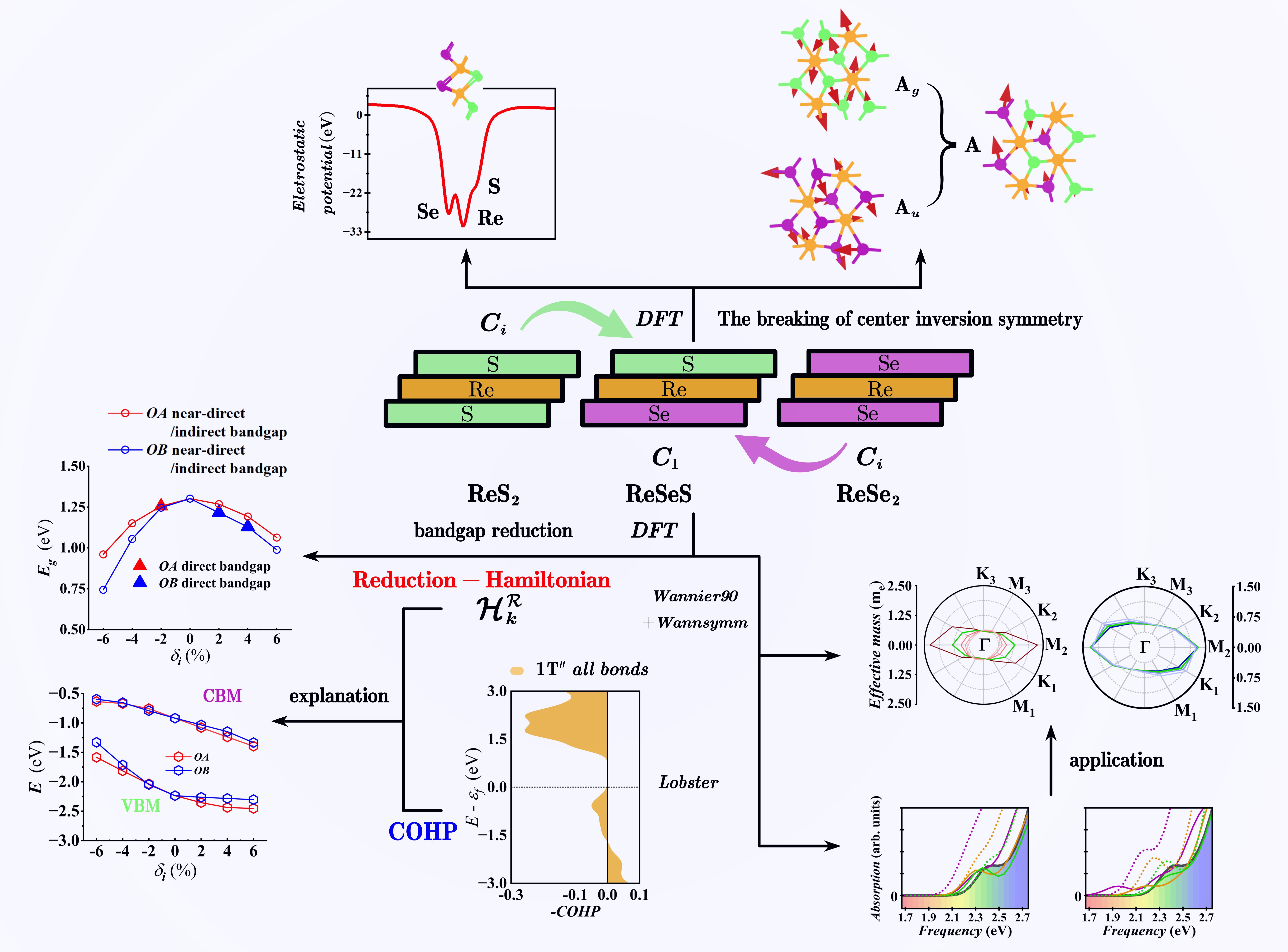}
\end{figure}
\end{graphicalabstract}

\begin{highlights}
\item The difference in antibonding strength between the valence and conduct bands increases the system's metallicity with both tensile and compressive deformations.
\item The large number of independent hoppings in the low-symmetry system can be greatly reduced by constructing the analytic Reduction-Hamiltonian.
\item TMDs monolayers with theoretically lowest symmetry obtained by reconstructing the electronic structures of ReX$_2$ with stronger anisotropy.
\end{highlights}


\end{frontmatter}


\section{\label{sec1}Introduction}
 Two-dimensional (2-D) materials always have a nanoscale thickness along with the length/width in the micrometer or even millimeter order of magnitude. They usually have excellent electrical, optical, and mechanical properties, making them have broad application prospects in electrical devices, optoelectronics, catalysis, energy storage, biomedical, sensors, flexible electronic devices, and other fields \cite{2,4,5,7}. Currently, there are several types of 2-D materials, which can be classified depending on their composition. One family of compounds known as two-dimensional transition metal dichalcogenides (TMDs) has received a lot of interest due to its unique physical features. Most TMDs are semiconductors, which gives them natural advantages in the production of transistors and other optoelectronic devices. What's more, their multilayer architectures also have a higher specific surface area, leading to a high potential for application in capacitor and sensor manufacture. \cite{8,10}.

The chemical formula of TMDs can be expressed as $\textrm{MX}_2$. M means transition metal atom, and X denotes the chalcogen. Based on the coordination of M, we can divide it into triangular prism coordination phase (like 2H or 3R phase) and octahedral coordination phase (like 1T or 2M phase) \cite{11,11-}. However, TMDs may experience more distortions before reaching a stable state. Hoffmann suggests that Peierls distortion reduces the symmetry of the $\textrm{ReSe}_2$ while minimizing its energy \cite{12,13}. The stable state is called the $\textrm{1T}^{\prime\prime}$ phase. It belongs to the space group $P\overline{1}$ and has only the center inversion operation. Many works indicate that TMDs will exhibit isotropic behaviors due to their symmetrical lattice structure. Similarly, when symmetry degrades, numerous anisotropic features will be induced \cite{14,16,17,18,19,20}. Liu’s team reports the anisotropy in $\textrm{ReSe}_2$ results in the ratio of renormalized field effect mobility as high as 3.1. Meanwhile, they also synthesized a digital inverter device from two Field-Effect-Transistor (FET) devices, demonstrating the application value of low symmetry TMDs \cite{21}.

However, the studies above have not achieved the minimum symmetry in TMD monolayers. In this work, we reconfigure the electronic structures of $\textrm{ReX}_2$ (M=S, Se) to obtain the ReSeS monolayer with Janus (double-face) structure. Because of the break in the center inversion symmetry, ReSeS belongs to the space group $P1$ and has no nontrivial point group operations. In theory, ReSeS's symmetry cannot be decreased anymore, and we shall analyze its electronic structures and regulation in Sec.~\ref{sec3} to demonstrate its anisotropic behaviors. We will compare the electronic structures and vibration spectrum between $\textrm{ReX}_2$ and ReSeS in Sec.~\ref{sec3a} to reflect the symmetry breaking. Meanwhile, the breaking of center inversion means that the infrared and Raman active modes will no longer be mutually exclusive, which can be utilized as a fingerprint to screen or identify this material. In Sec.~\ref{sec3b}, We provide two approaches to understanding the bandgap decrease mechanism with geometric distortions. The Reduction-Hamiltonian provides explanations qualitatively, while the analyses of bonding/antibonding near the Fermi surface give more concise physical images. Finally, the effective mass and absorption spectrum of electrons regulated by the geometric deformations along the different directions are discussed in Sec.~\ref{sec3c} to demonstrate the potential application of the ReSeS monolayer.
\section{\label{sec2}Methods}
\subsection{\label{sec2a}Density functional theory (DFT) method}
All the theoretical calculations are based on the Density functional theory (DFT) \cite{22,23,24}, using the Perdew-Burke-Ernzerh of generalized gradient approximation (GGA-PBE) \cite{26} as exchange-correlation function. We utilize the CASTEP \cite{28} and VASP \cite{29,30} code, and the vacuum space in our models is set to more than 15 $\textrm{\AA}$ to prevent the interaction between periodic layers. For the CASTEP code with OTFG norm conserving pseudo-potential, the convergence tolerance of the self-consistent cycle is set as $1.0\times10^{-7}$ eV. And it is set as $1.0\times10^{-5}$ eV for VASP code using projector augmented wave (PAW) pseudo-potential. Table~\ref{table1} lists the cut-off energy and Monkhorst-Pack k-point meshes meeting the convergence requirements of some structures. Wherein, the $\textrm{1T}^{\prime}$ phase is a metallic metastable phase, and the phase transition from it to the stable $\textrm{1T}^{\prime\prime}$ phase will be investigated in the Sec.~\ref{sec3b2}. 

In this work, we alter the structures' lattice parameters to imitate the material's geometrical deformation happening in the real world. In calculations, this is achieved by manually adjusting the relevant lattice parameters and optimizing only the atomic coordinates. It is worth emphasizing that although we only modify the lattice parameter in one direction at a time, this is distinct from uniaxial strain in the literal sense of the term. Because we do not automatically optimize the lattice parameters in one axis while keeping the lattice parameters fixed in the other. In other words, the geometric deformations applied are a kind of anisotropic biaxial strain. Seeing parameter settings for geometry optimization in \textit{Supplementary 1}.
\begin{table}
\centering
\renewcommand\arraystretch{1.3}
\footnotesize
\begin{tabular}{cccc}
\hline
&\textrm{Codes}&\textrm{Cut-off}&\\
\textrm{Structure}&\textrm{used for}&\textrm{energy}&\textrm{k-points}\\
&\textrm{calculation}&\textrm{(eV)}&($x\times{y}\times{z}$)\\
\hline
$\textrm{1T}^{\prime\prime}-\textrm{ReSeS}$ &\textrm{VASP}& 500 & $9\times9\times1$\\
$\textrm{1T}^{\prime}-\textrm{ReSeS}$&\textrm{VASP}& 500 & $18\times9\times1$\\
$\textrm{1T}^{\prime\prime}-\textrm{ReSeS}$ &\textrm{CASTEP}& 1450 & $4\times4\times1$\\
$\textrm{1T}^{\prime}-\textrm{ReSeS}$ &\textrm{CASTEP}& 1550 & $11\times4\times1$\\
$\textrm{ReS}_2$  &\textrm{CASTEP}& 900&$4\times4\times1$\\
$\textrm{ReSe}_2$  &\textrm{CASTEP}& 1500&$4\times4\times1$\\
\hline
\end{tabular}
\caption{\label{table1}The cut-off energy and Monkhorst-Pack k-points meshes of some structures.}
\end{table}
\subsection{\label{sec2b}Anlysis tools}
Based on the DFT results, certain analysis techniques are employed for further investigations. In Sec.~\ref{sec3b1}, we use the Wannier90 \cite{31,32,33} and WannSymm \cite{35} codes to construct the Reduction-Hamiltonian. Wannier90 presents a straightforward method for obtaining the tight-binding-like Hamiltonian, and WannSymm guarantees its symmetry. In Sec.~\ref{sec3b2}, Lobster \cite{36} codes are used to show the bonding/antibonding properties of the electronic structures near the Fermi surface. As a potent method utilized extensively in covalent bond analyses, it provides great chemical intuition by calculating the crystal orbital Hamilton population (COHP). Based on its findings, we present simple and straightforward physical processes for the decrease of bandgaps under geometric deformations regulation.

\begin{figure}
\centering
\includegraphics[width=14cm]{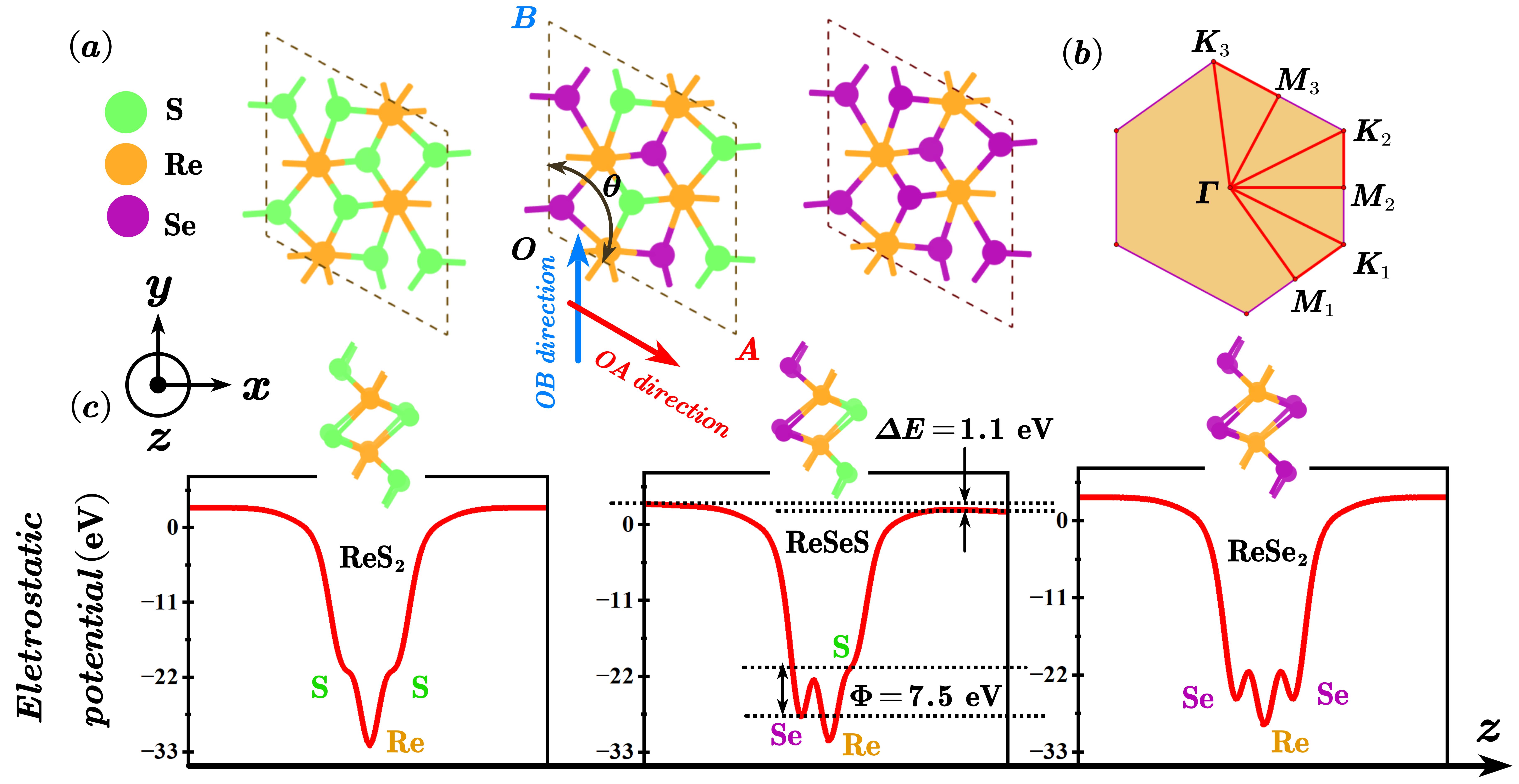}
\caption{\label{fig1}(a) Top view of the unit cell and (b) first Brillouin zone of $\textrm{ReS}_2$, $\textrm{ReSeS}$, and $\textrm{ReSe}_2$ monolayer. The red and blue arrows in (a) represent the OA and OB directions, respectively. Red lines in (b) represent the k-path in reciprocal space. (c) Eletrostatic potential of $\textrm{ReS}_2$, $\textrm{ReSeS}$, and $\textrm{ReSe}_2$ monolayer along the $z$ axis.}
\end{figure}
\section{\label{sec3}Results and discussion}
\subsection{\label{sec3a}Structures and symmetry}
ReSeS is one of the TMDs materials with a distinct layer structure in the bulk, where the layers are joined by weak intermolecular forces (Van Der Waals interaction), allowing the ReSeS monolayer to be separated from the bulk to form a 2-D material. ReSeS monolayer retains three atom layers, with the Se/S-layer positioned on the two sides, demonstrating that ReSeS has a Janus structure with lower symmetry than $\textrm{ReX}_2$. The ReSeS's unit cell has 12 non-identical atoms, and its stable structure belongs to the space group $P1$, which means there are no nontrivial point group operations. Furthermore, the metal atoms in ReSeS are 8-coordinated, and it belongs to the $\textrm{1T}$ class in the TMD family. Similar to $\textrm{ReX}_2$, its stable structural lattice experiences pronounced distortions and is named as $\textrm{1T}^{\prime\prime}$ phase. Fig.~\hyperref[fig1]{1(a)} displays the lattice structures of $\textrm{ReS}_2$, $\textrm{ReSeS}$, and $\textrm{ReSe}_2$ monolayer, respectively. Their lattice parameters are presented in Table~\ref{table2}. The lattice vector $\textbf{\textit{OA}}$ and $\textbf{\textit{OB}}$ grow as the number of Se atoms in the compound increases due to the longer covalent bond lengths between Re and Se. Fig.~\hyperref[fig1]{1(b)} depicts the first  Brillouin zone (1-BZ) of this sort of TMDs. Because of the proximity of the lattice vector $\textbf{\textit{OA}}$ and $\textbf{\textit{OB}}$, similar to the graphene, its 1-BZ is nearly orthogonal. However, the lower symmetry and anisotropic structure lead to three unequal sets of M/K points in 1-BZ. Along with the $\Gamma$ point, they comprise the irreducible k-path in reciprocal space.

Because of the large difference in the outer electron distribution and nuclear charge between Se and S, the Helmholtz free energy $\mathcal{F}$ of ReX$_2$ tends to increase with the proportion of Se. Under the pseudo-potential adopted, the ReSeS's energy is lower than that of ReS$_2$ (11744.3 eV), which is very similar to the energy of ReSeS over ReSe$_2$ (11744.7 eV), and also close to the 4 times difference in the energies of the isolated Se (-3236.3 eV) and S (-299.2 eV) atoms. These illustrate the accuracy of our calculations as well as the stability brought about by the doping of Se atoms.

When comparing the $\textrm{ReX}_2$ and ReSeS structures, the double-face feature breaks ReSeS's center inversion symmetry. It also means that electronic structures are no longer symmetrical on either side of the Re-layer, and can be described directly by electrostatic potential. As seen in Fig.~\hyperref[fig1]{1(c)}, the electrostatic potential has some minimal values where the ions are dispersed, indicating that the electrons tend to be stabilized in these areas. Unlike $\textrm{ReX}_2$'s symmetric electrostatic potential, a potential energy difference of $\Phi=7.5$ eV exists between the Se-layer and S-layer due to the symmetry breaking. Because Se has a higher nuclear charge than that of S, the electrons distributed around the Se-layer have lower energy, forming a structure similar to a one-dimensional potential well along the $z$ axis. Work functions also embody the difference in electron distribution. Due to the stronger binding of the Se-layer to electrons, its value on the Se-layer side is approximately $\Delta E=1.1$ eV greater than that on the S-layer side. $\Delta E$ reflects the magnitude of intrinsic dipole induced by the asymmetric Janus structure and is larger than that reported in other selenium-sulfur metallic compounds, such as 0.7 eV in WSeS monolayer \cite{40,41}. Therefore, the large dipole gives ReSeS great advantages in n/p-type diode design, highlighting the potential for its application \cite{42}. 
\begin{table}[t]
\centering
\renewcommand\arraystretch{1.3}
\footnotesize
\begin{tabular}{cccc}
\hline
Structures&OA ($\textrm{\AA}$)&OB ($\textrm{\AA}$)&
$\theta$ ($^{\circ}$)\\
\hline
$\textrm{ReS}_2$&6.51&6.40&118.85\\
$\textrm{ReSeS}$&6.65&6.53&118.84\\
$\textrm{ReSe}_2$&6.78&6.66&118.82\\
\hline
\end{tabular}
\caption{\label{table2}Lattice parameters of $\textrm{ReS}_2$, $\textrm{ReSeS}$, and $\textrm{ReSe}_2$ monolayer.}
\end{table}
\begin{table}
\centering
\renewcommand\arraystretch{1.3}
\setlength\tabcolsep{3pt}
\footnotesize
\begin{tabular}{cccccccccccc}
\hline
\multicolumn{4}{c}{$\textrm{ReS}_2$}&\multicolumn{4}{c}{$\textrm{ReSeS}$}&\multicolumn{4}{c}{$\textrm{ReSe}_2$}\\
\cline{1-4}\cline{5-8}\cline{9-12}
Fre.&I.r.&Fre.&I.r.&Fre.&I.r.&Fre.&I.r.&Fre.&I.r.&Fre.&I.r.\\
(cm$^{-1}$)&&(cm$^{-1}$)&&(cm$^{-1}$)&&(cm$^{-1}$)&&(cm$^{-1}$)&&(cm$^{-1}$)&\\
\hline
130.3&$\textrm{A}_g$&299.1&$\textrm{A}_g$&111.0&A&229.3&A&100.9&$\textrm{A}_g$&194.3&$\textrm{A}_g$\\
135.0&$\textrm{A}_u$&300.9&$\textrm{A}_u$&117.6&A&240.7&A&114.0&$\textrm{A}_u$&198.6&$\textrm{A}_u$\\
137.8&$\textrm{A}_g$&304.3&$\textrm{A}_g$&119.4&A&252.1&A&116.7&$\textrm{A}_g$&201.6&$\textrm{A}_g$\\
148.9&$\textrm{A}_g$&311.9&$\textrm{A}_g$&123.4&A&254.2&A&117.5&$\textrm{A}_g$&207.4&$\textrm{A}_u$\\
154.6&$\textrm{A}_u$&312.8&$\textrm{A}_u$&129.6&A&265.0&A&119.2&$\textrm{A}_u$&209.3&$\textrm{A}_g$\\
159.4&$\textrm{A}_g$&332.2&$\textrm{A}_u$&141.3&A&271.4&A&123.9&$\textrm{A}_g$&215.3&$\textrm{A}_u$\\
210.2&$\textrm{A}_g$&332.9&$\textrm{A}_g$&164.0&A&278.2&A&139.1&$\textrm{A}_g$&228.4&$\textrm{A}_g$\\
216.5&$\textrm{A}_u$&351.5&$\textrm{A}_u$&169.9&A&283.6&A&144.1&$\textrm{A}_u$&228.9&$\textrm{A}_u$\\
230.7&$\textrm{A}_g$&355.0&$\textrm{A}_g$&185.5&A&292.8&A&166.5&$\textrm{A}_g$&231.7&$\textrm{A}_g$\\
259.7&$\textrm{A}_u$&362.5&$\textrm{A}_u$&191.6&A&300.6&A&171.6&$\textrm{A}_u$&242.1&$\textrm{A}_u$\\
262.9&$\textrm{A}_g$&363.1&$\textrm{A}_g$&195.9&A&301.5&A&172.8&$\textrm{A}_g$&244.4&$\textrm{A}_g$\\
263.2&$\textrm{A}_u$&383.4&$\textrm{A}_u$&199.2&A&317.7&A&173.5&$\textrm{A}_u$&245.2&$\textrm{A}_u$\\
269.6&$\textrm{A}_g$&393.4&$\textrm{A}_g$&206.7&A&338.9&A&180.9&$\textrm{A}_g$&260.3&$\textrm{A}_g$\\
275.6&$\textrm{A}_u$&406.2&$\textrm{A}_g$&208.2&A&350.4&A&184.7&$\textrm{A}_u$&296.2&$\textrm{A}_g$\\
296.7&$\textrm{A}_g$&411.1&$\textrm{A}_u$&223.1&A&375.2&A&188.2&$\textrm{A}_g$&300.4&$\textrm{A}_u$\\
297.1&$\textrm{A}_u$&424.3&$\textrm{A}_g$&224.3&A&397.4&A&191.5&$\textrm{A}_u$&303.8&$\textrm{A}_g$\\
&&443.6&$\textrm{A}_u$&&&442.9&A&&&320.9&$\textrm{A}_u$\\
\hline
\end{tabular}
\caption{\label{table3}Frequencies (Fre.) and irreducible representations (I.r.) (except for 3 acoustic branches) corresponding to the vibration modes of $\textrm{ReX}_2$ and ReSeS monolayer.}
\end{table}

Symmetry breaking can also alter the infrared absorption and Raman spectrum. In general, if the vibration of ions changes the dipole moment of the system, the associated vibration mode may be infrared active. A vibration mode that alters the system's polarisation rate could be Raman active. Group theory demonstrates that in a center-symmetric system, the infrared active modes have a mutually exclusive connection to Raman active modes. $\textrm{ReX}_2$ belong to the point group $C_i$, which contains the center inversion operation. Hence excluding the 3 acoustic branches representing the overall motion, 15 asymmetric $\textrm{A}_u$ modes are infrared active, and 18 symmetric $\textrm{A}_g$ modes are Raman active. In contrast, the symmetry-free ReSeS belongs to the point group $C_1$ and has just one kind of mode A, and all 33 modes are both infrared active and Raman active. This is an inherent characteristic owing to its Janus structure, and all these modes are listed in Table~\ref{table3}. 

\begin{figure}[!t]
\centering
\includegraphics[width=14cm]{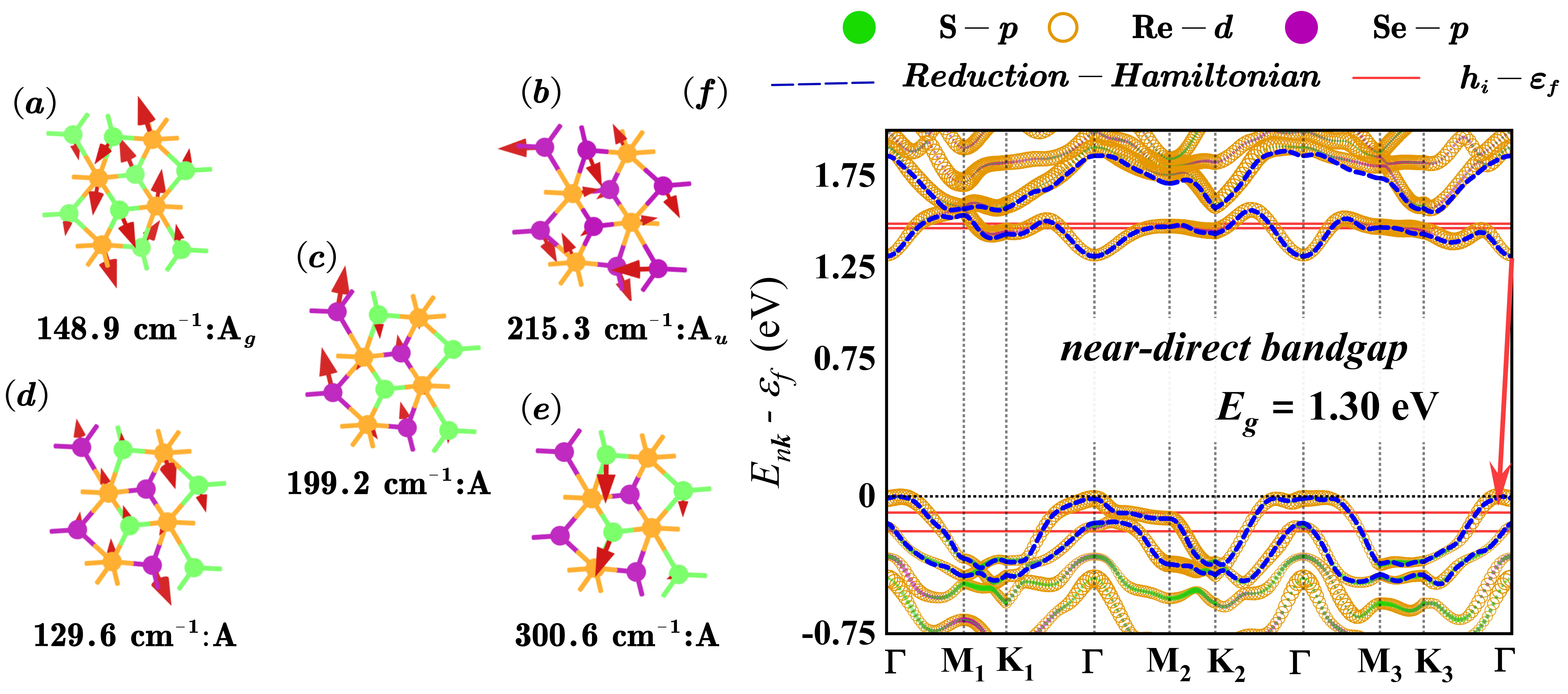}
\caption{\label{fig2}(a)-(e) Some typical vibration modes with frequencies and irreducible representations of $\textrm{ReX}_2$ and ReSeS monolayer. The red arrows represent the directions of the ions' vibrations, and the larger size reflects the larger magnitude. (f) The band structure of the ReSeS monolayer. Green dots, yellow circles, and purple dots represent the atomic orbital weight on the corresponding Bloch state's projection. The blue dashed curve displays the calculation results of the Reduction-Hamiltonian, while the red solid straight line displays the difference between the energy of the mix-orbital and the Fermi energy. The red arrow points from CBM to VBM.}
\end{figure}

Some typical vibration modes of $\textrm{ReX}_2$ and ReSeS monolayer are displayed in Fig.~\hyperref[fig2]{2} (symmetric Fig.~\hyperref[fig2]{2(a)} and asymmetric Fig.~\hyperref[fig2]{2(b)} vibrations). It can also be discovered that in the low-frequency area, Re, Se, and S all vibrate, however as the frequency increases, only the S can remain vibrating. This can be simply understood through the classical harmonic oscillation, where the frequency of vibration $\nu$ is proportional to the $m^{-1/2}$, and atoms with light mass $m$ are more likely to vibrate at high frequencies. On the one hand, metal atoms in TMDs always have higher mass, therefore their vibrations mostly produce low-energy phonons; on the other hand, due to the larger mass of Se, the frequencies of all modes will demonstrate a decreasing trend as the proportion of Se in the system increases, which is also consistent with the results in Table~\ref{table3}. Generally, the ReSeS monolayer exhibits 33 infrared/Raman active modes, which is more than most conventional 2-D materials (save for particular doping) and provides a wide spectrum observation interval \cite{46,47,49}. Its distinct vibration spectrum fingerprints also offer effective guidance for material identification.
\subsection{\label{sec3b}Electronic structures and regulation}
Stable $\textrm{1T}^{\prime\prime}-$ReSeS present semiconductivity, and Fig.~\hyperref[fig2]{2(f)} illustrates it has a wide bandgap $E_g=1.30$ eV. Three unequal triangular loops can be found in the k-path and each can be roughly represented as the irreducible Brillouin zone of hexagonal lattice (The band in Fig.~\hyperref[fig2]{2(f)} is calculated with PAW pseudo-potential. And it can be proved to be reliable by comparing the results obtained from OTFG norm conserving pseudo-potential in \textit{Supplementary 1}). Its maximum of valence band (VBM) is not exactly located at the high symmetry $\Gamma$ point. However, we also define the bandgap type as near-direct bandgap as its minimum of conduct band (CBM) is located at the $\Gamma$ point precisely. Similar effects are noted in $\textrm{ReX}_2$. And when the fraction of Se atoms in the structure rises, the bandgap tends to decrease, according to a comparison of the bandgap sizes of the three configurations (1.43 eV and 1.21 eV for $\textrm{ReS}_2$ and $\textrm{ReSe}_2$'s bandgap, respectively. Seeing their band structures in \textit{Supplementary 1}). This implies that Se-doping might be able to increase the system's metallicity, which is supported by related research \cite{50}. More discussion related to doping is in \textit{Supplementary 3}.

\begin{table*}[!t]
\centering
\caption{\label{table4}Energy of V/CBM calculated from work function of $\textrm{ReS}_2$, $\textrm{ReSeS}$, and $\textrm{ReSe}_2$ monolayer. For the calculation of the Average, we just take the average of the difference between the Fermi energy and the vacuum energy of the Se/S-layer side.}
\renewcommand\arraystretch{1.3}
\footnotesize
\begin{tabular}{cccccc}
\hline
&$\textrm{ReS}_2$&\multicolumn{3}{c}{ReSeS}&$\textrm{ReSe}_2$\\
\cline{3-5}
&&S-layer&Se-layer&Average&\\
\hline
$E_{VBM}=\varepsilon_f-E_{vac}$ (eV)&-5.68&-4.79&-5.87&-5.33&-4.89\\
$E_{CBM}=\varepsilon_f-E_{vac}+E_g$ (eV)&-4.25&-3.49&-4.57&-4.03&-3.68\\
\hline
\end{tabular}
\end{table*}

The difference between the vacuum $E_{vac}$ and the Fermi energy $\varepsilon_f$ can be represented easily by the work function. Considering the VBM and the Fermi energy have similar physical meaning at 0 temperature for semiconductors, the difference between the two can be regarded as both the work function and the VBM approximately. And the CBM can then be derived from the relation $\varepsilon_f-E_{vac}+E_g$. Results are listed in the Table~\ref{table4}. Since Se has stronger binding for electrons (note that these electrons cannot be interpreted as the outside electrons by default) as mentioned in Sec.~\ref{sec3a}, the $E_{VBM}$ obtained from the Se-layer side will be smaller, which is also consist with the statistics displayed in the Table~\ref{table4}. Meanwhile, outer electrons of Se have higher energy, which means that the electrons in $\textrm{ReSe}_2$ will occupy higher energy states, resulting in a larger energy of V/CBM than the $\textrm{ReS}_2$. Combining band structures of $\textrm{ReX}_2$ with data listed in Table~\ref{table4} and observing the change of V/CBM, it is found that the dispersion relation inside the valence and conduct band is similar, but as the V/CBM in $\textrm{ReSe}_2$ rises, VBM increases more rapidly, leading to the narrowing of the difference between the C/VBM and a smaller bandgap. This implies that a more fundamental reason for the decreasing of $\textrm{ReSe}_2$'s bandgap is not the change in the dispersion relation of a specific band, but rather the difference in the amount of changes between distinct bands. This variation process of bandgap is a significant specificity of these materials, and as will be seen later in Sec.~\ref{sec3b2}, it also explains the change in bandgaps caused by the geometric deformations regulation.

In real situations, two-dimensional materials will constantly experience a variety of deformations. Correspondingly, geometric distortions will cause changes in the electronic structures, leading to variations in the properties. Strain engineering \cite{51} is frequently utilized to control the electrical and optical properties of 2-D materials. In this work, we defined the deformation factor $\delta _\textbf{\textit{i}}=(l_{\textbf{\textit{i}}}^\prime-l_{\textbf{\textit{i}}})/l_{\textbf{\textit{i}}}$ to show the distortions. The $\textbf{\textit{i}}$ indicates the direction of deformation, and $l_{\textbf{\textit{i}}}^\prime$/$l_{\textbf{\textit{i}}}$ represent the lattice parameter after/before the deformation. It should be firstly noted that all the distorted structures we investigated, according to the phonon spectra in \textit{Supplementary 1}, are within the elastic limit of ReSeS monolayer. Our calculations reveal that geometric deformations along the OA or OB direction, whether tensile or compressive, cause the bandgap of the ReSeS monolayer to decrease. Fig.~\hyperref[fig4]{3(a)} displays the results, and similar phenomena have also been found in other TMDs \cite{52,52p,53}. For ReSeS, except for cases $\delta_{\bm{OA}}=-2\%$, $\delta_{\bm{OB}}=2\%$, and $\delta_{\bm{OB}}=4\%$ where the bandgap exhibits as direct bandgap, the bandgap in the remaining cases is either near-direct or indirect. 

Despite ample observations, there is still a dearth of investigation into the mechanism of the decrease of bandgaps. Some studies claim that raising pressure can effectively reduce the material's bandgap, which is similar to our discussions of compressive deformation. This can be explained by some TMDs' negative pressure coefficient \cite{54,55}. The problem, however, is that this is difficult to generalize to the case of stretching or tensing, so in general, this method only explains half of this phenomenon. Furthermore, pressurization might mainly focus on modifying the interlayer interaction, which is likewise not applicable for ReSeS monolayers; Others argue the compressive strain shortens the bonds and strengthens the hopping between atoms, resulting in a broadened bandwidth and a narrow bandgap. This explanation is more intuitive but still does not allow for the situations of tensile and compressive deformation at the same time \cite{19}. To answer this question, we provide two ideas in Sec.~\ref{sec3b1} and Sec.~\ref{sec3b2} to examine the roots of this phenomenon.

The bandgap size is determined by C/VBM alone mathematically, and the first thing to consider before delving into the specifics is how C/VBM varies mainly. The calculation results of C/VBM are displayed in Fig.~\hyperref[fig4]{3(b)}. Energy changes in the band's specific dispersion shape, as well as for the entire band, can result in variations in C/VBM. As the typical band structures are shown in Fig.~\hyperref[fig4]{3(c), (d)}, the changes in the dispersion shape of bands are due to the increase in the valence band and drop in conduct band happening not at the $\Gamma$ point under geometric deformations $\delta_{\bm{OA}}=6\%$ and $\delta_{\bm{OB}}=-6\%$, respectively. These both lead to an undulation in the band's energy of about $\delta E=0.07$ eV. In fact, most $\delta E$ in the calculations are distributed between 0.05 eV and 0.1 eV, and occasionally up to 0.2 eV, which is enough to lead to transformations of bandgap's type (direct, near-direct, and indirect bandgap), but is still smaller than the magnitude of the variation of C/VBM in Fig.~\hyperref[fig4]{3(b)} and bandgap narrowing. Therefore, compared to the change in dispersion shape of the valence or conduct band, the overall energy change of bands is a more essential factor responsible for the decrease of bandgaps. And the following discussions will focus on its mechanism.
\subsubsection{\label{sec3b1}Reduction-Hamiltonian}
The ReSeS monolayer's tight-binding Hamiltonian may be reduced to a Reduction-Hamiltonian $\hat{H}^\mathcal{R}$ by creating 4 isolated mix-orbitals. The $\hat{H}^\mathcal{R}$ solely considers the interaction between these mix-orbitals, obtaining 4 bands near the Fermi surface. And its kernel $\mathcal{H}_{\bm{k}}^{\mathcal{R}}$ is provided by Eq.~\ref{eq1}.
\begin{equation}
\mathcal{H}_{\bm{k}}^{\mathcal{R}} = \bigoplus_{i=1}^4h_i+\Delta_{\bm{k}}
\label{eq1}
\end{equation}

Where $h_i$ is the kernel of single mix-orbital Hamiltonian, whereas $\Delta_{\bm{k}}$ indicates the interaction between them. Both $\mathcal{H}_{\bm{k}}^{\mathcal{R}}$ and $\Delta_{\bm{k}}$ are $4\times 4$ hermitian matrices. Due to the time reversal symmetry of the ReSeS monolayer, all 4 $h_i$ are real numbers. The derivation of this tight-binding-like Hamiltonian can be seen in \textit{Supplementary 2}.

\begin{figure}[!b]
\centering
\includegraphics[width=7cm]{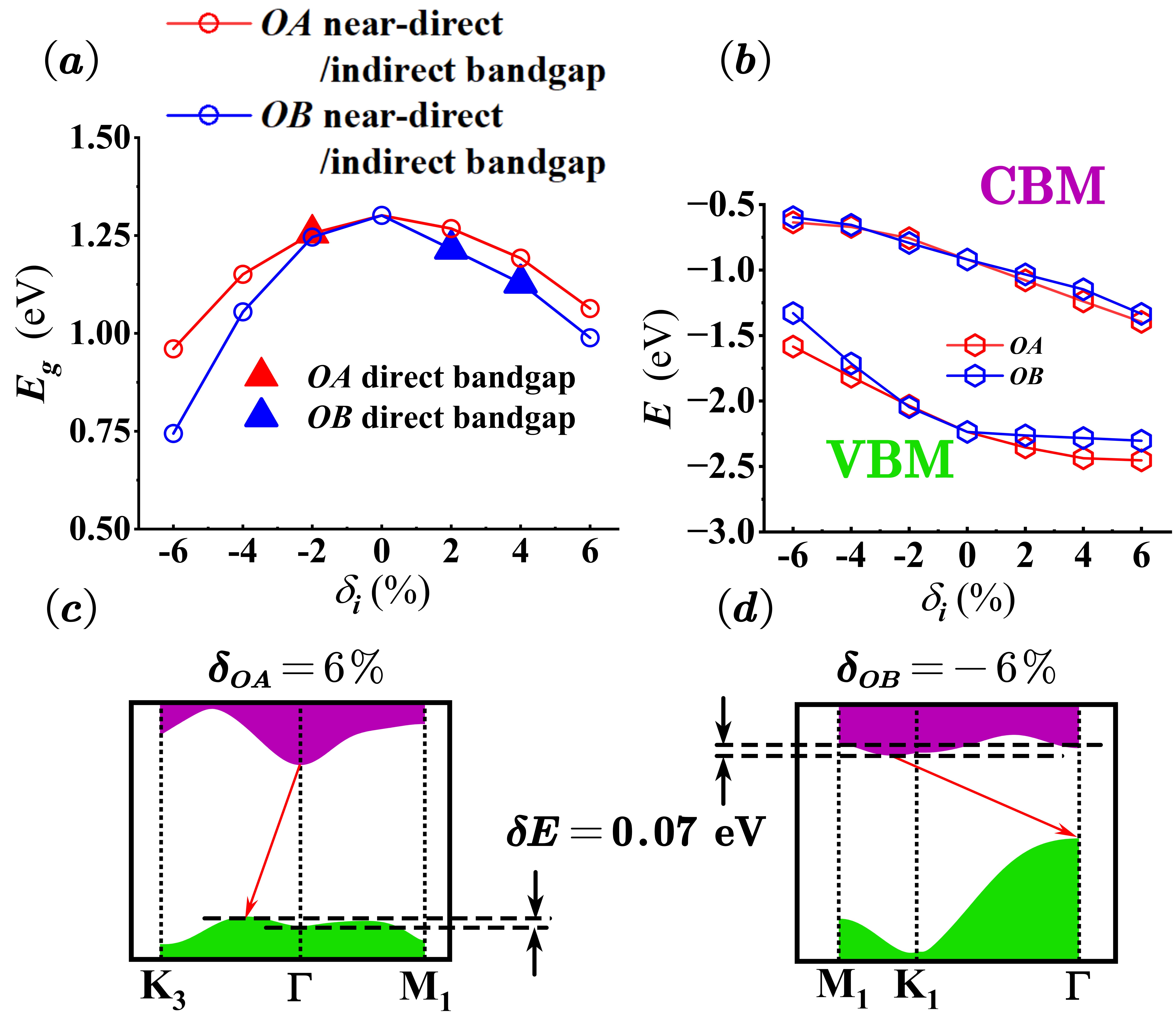}
\caption{\label{fig4}(a) Bandgap's and (b) C/VBM's changes under geometric deformations regulation. Red/blue lines represent the geometric deformations along the OA/OB direction, respectively. Hollow circle and solid triangle in (a) means near-direct/indirect bandgap and direct bandgap, respectively. (c), (d) Schematic diagrams of the bandgaps under two geometrical deformations ($\delta_\textbf{\textit{OA}}=6\%$ and $\delta_\textbf{\textit{OB}}=-6\%$). The purple/green region represents the conduct/valence band, and the red arrow points from CBM to VBM. $\delta E$ means energy changes induced by the variation of the dispersion shape of the valence or conduct band.}
\end{figure}
The energy level is typically discretized in isolated systems, such as molecules placed in a vacuum. In contrast, for solid systems, space translation symmetry enables the creation of a continuous energy spectrum in reciprocal space. This process can be seen as the interaction of isolated energy levels causing the split, which results in the formation of bands with specific spreading widths. In contrast, the band's energy can also represent the energy of the associated level. For this Reduction-Hamiltonian, $h_i$ represents the energy levels of 4 isolated mix-orbitals, which do not contain any dispersion information but form 4 bands located near the Fermi surface with full dispersion relations by the interaction $\Delta_{\bm{k}}$ between them. Both the energy of 4 mix-orbitals and band structures calculated from this Hamiltonian are displayed in Fig.~\hyperref[fig2]{2(f)}. This model appears to be in good agreement with the DFT results, and the energy range of $h_i-\varepsilon_f$ also corresponds to the four bands near the Fermi surface. These reflect its reliability. 

\begin{figure}[!t]
\centering
\includegraphics[width=7cm]{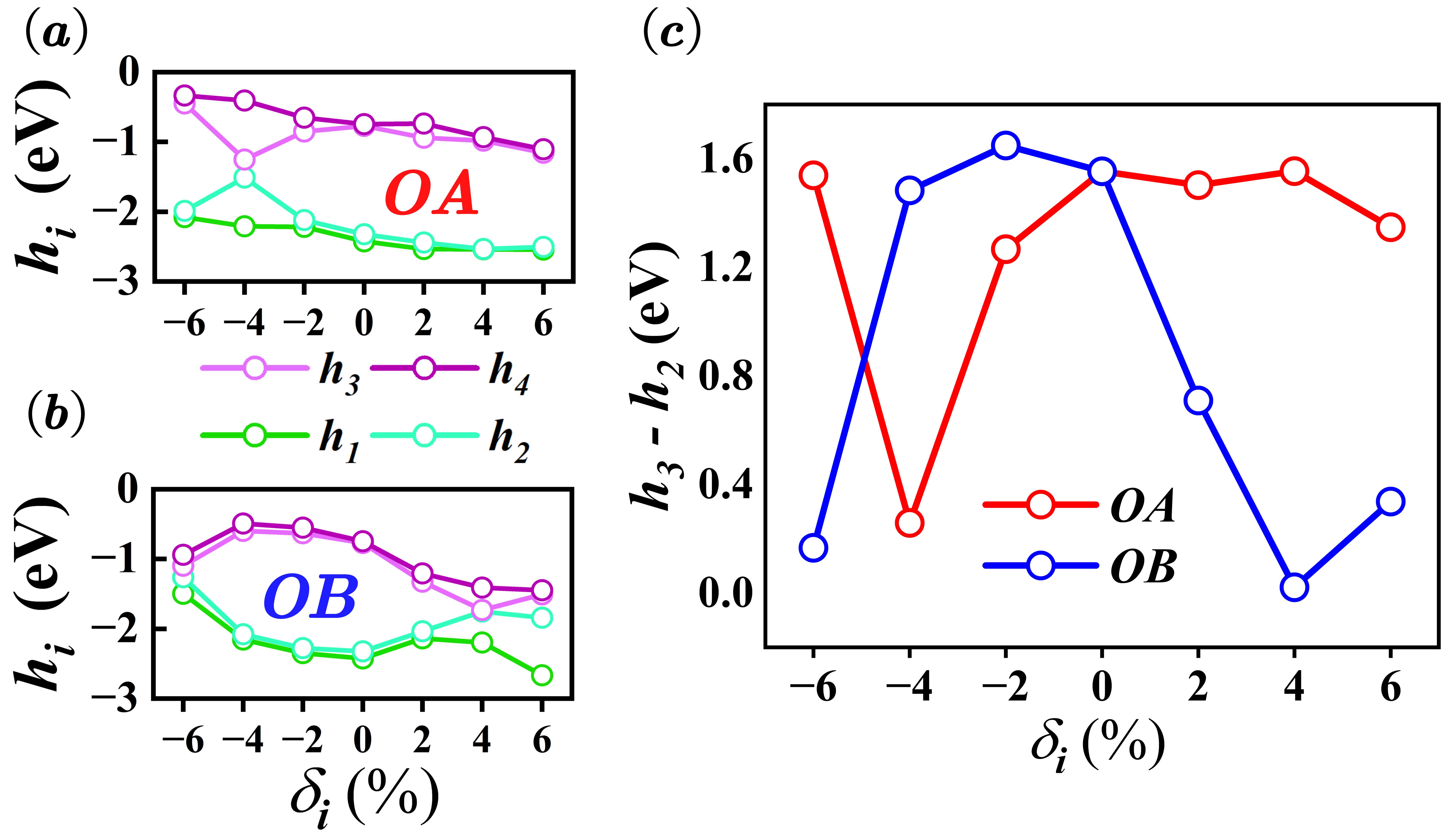}
\caption{\label{fig5}The variations of $h_i$ under geometric deformations along the (a) OA direction and (b) OB direction. (c) the variation of $h_3-h_2$ under geometric deformations along the OA and OB direction.}
\end{figure}

Our calculation results indicate that the energy of $h_i$ is consistently larger than that of interaction described by $\Delta_{\bm{k}}$ approximately. Therefore, we can simply assume that the variation of $h_2$ and $h_3$ represents the overall energy change of the valence band and conduct band, and their difference $h_3-h_2$ reflects the value of $E_g$. The value of $h_i$ along with the $h_3-h_2$ under geometric deformations regulation are shown in Fig.~\hyperref[fig5]{4}, and more detail about Reduction-Hamiltonian results with distorted structures can also be found in \textit{Supplementary 2}. In Fig.~\hyperref[fig5]{4(a), (b)}, the energy of all $h_i$ decreases as the geometric deformations change from compressive to tensile, which is following the variations of C/VBM in Fig.~\hyperref[fig4]{3(b)}, while $h_3-h_2$ in Fig.~\hyperref[fig5]{4(c)} can also be roughly seen as decreasing with the increase of the geometric deformations. As a result, in our model, the reduction of the energy difference between the isolated mix-orbitals increases the metallicity of the ReSeS monolayer under geometric deformations regulation.

The essence of $h_i$ is the hopping of the Wannier function itself. Independent hoppings are few in systems with high symmetry, resulting in a concise analytic description of the tight-binding Hamiltonian \cite{56,57}. Unfortunately, the lack of symmetry restrictions in the ReSeS monolayer leads to the number of tight-binding Wannier functions constructed from bands near the Fermi surface being as high as 44 (20 corresponding to the $d$ orbital of Re, 12 corresponding to the $p$ orbital of Se, and 12 corresponding to the $p$ orbital of S). Indeed, we can artificially reduce the number of Wannier functions in the calculation to simplify the Hamiltonian, where each Wannier function is equal to the combination of some atomic orbitals, i.e., mix-orbital. However, mixing more orbitals leads to worse localization and symmetry when utilizing the MLWF method. Nonetheless, ReSeS can still obtain a Reduction-Hamiltonian having just 4 mix-orbitals since its symmetry-free structure minimizes the impact of poor symmetry of Hamiltonian. This also gives suggestions for theoretical research into other low-symmetry TMDs.

Overall, the Reduction-Hamiltonian gives a straightforward explanation for bandgap reduction by creating 4 mix-orbitals. However, it is unable to account for either the reduction of $h_i$ with tensile deformations enhancement or the reduction of $h_3-h_2$ with geometric distortions. Worse, certain results do not accord with the variations of the bandgap very well (such as $h_3-h_2$ at $\delta_{\bm{OA}}=-6\%$). Therefore, Reduction-Hamiltonian just provides another way to understand the phenomena and enables basic qualitative analyses. And the investigations in Sec.~\ref{sec3b2} will solve this problem.
\subsubsection{\label{sec3b2}Bonding/antibonding properties near the fermi surface}
\begin{figure}
\centering
\includegraphics[width=14cm]{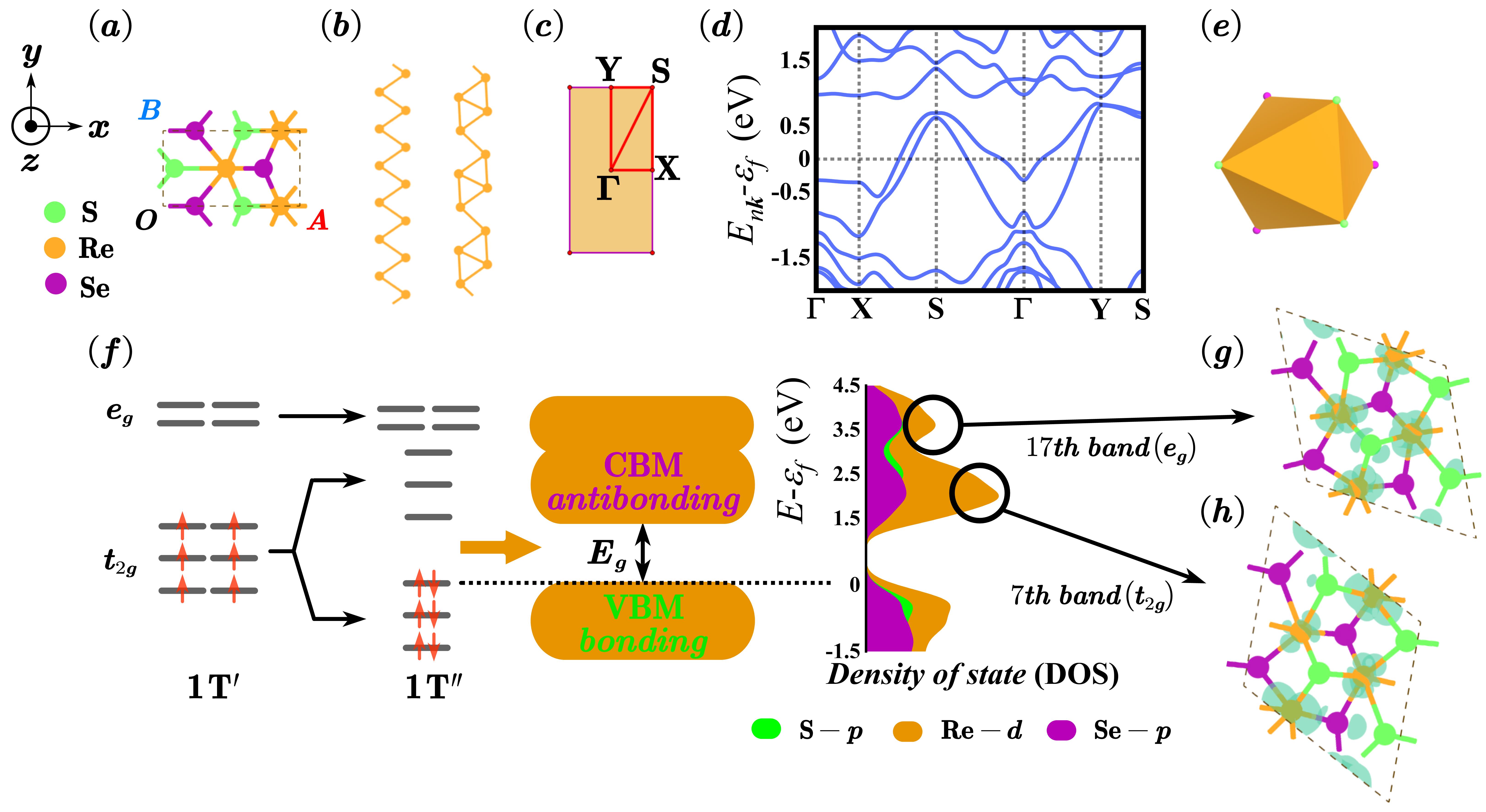}
\caption{\label{fig6}(a) Structures of 1T$^{\prime}-\textrm{ReSeS}$ monolayer. (b) Re-chains of 1T$^{\prime}-\textrm{ReSeS}$ (left) and 1T$^{\prime\prime}-\textrm{ReSeS}$ (right) monolayer. (c) First Brillouin zone and (d) band structure of 1T$^{\prime}-\textrm{ReSeS}$ monolayer. The red lines in (c) represent the k-path in reciprocal space. (e) Distorted octahedral coordination of Re. (f) Schematic diagram of electron filling during the phase transition from the 1T$^{\prime}$ to 1T$^{\prime\prime}$, and the density of state of 1T$^{\prime\prime}-\textrm{ReSeS}$ monolayer. The yellow arrow and region on the left side mean isolated energy levels split into band structures with dispersion relations. The green, yellow, and purple regions in the density of states represent the contributions of S's $p$ orbitals, Re's $d$ orbitals, and Se's $p$ orbitals, respectively. (g), (h) Distribution of electronic orbitals corresponding to the bands in 1T$^{\prime\prime}-\textrm{ReSeS}$ monolayer, The number represents the ordinal numbers of the band, and $e_g$ and $t_{2g}$ indicates the types of Re's $d$ orbitals. The black circles and arrows mean the structures with $d$ orbitals and the energy range of their corresponding bands in the density of states.}
\end{figure}
To explain the phenomena of bandgap reduction in ReSeS, we need first to clarify the mechanism of the bandgap creation. Compared to the 1T phase, which possesses a hexagonal lattice, the 1T$^{\prime\prime}$ phase undergoes significant distortions. For ReX$_2$, there is a metastable metallic phase called 1T$^{\prime}$ phase during the phase transformation to the 1T$^{\prime\prime}$ \cite{58}, which is crucial for understanding bandgap's formation. Similarly, ReSeS monolayer also has 1T$^{\prime}$ phase, and its structure are displayed in Fig.~\hyperref[fig6]{5(a)}. We calculate the phonon dispersion of 1T$^{\prime\prime}$ and 1T$^{\prime}$ phases to clarify their stability, and the results are provided in \textit{Supplementary 1}. The main feature of the phase transition from 1T$^{\prime}$ to 1T$^{\prime\prime}$ is the 4-polymerization effect that happens in Re-chain \cite{12}, which converts the zigzag-chain in the 1T$^\prime$ phase to the diamond-chain in the 1T$^{\prime\prime}$ phase, as shown in Fig.~\hyperref[fig6]{5(b)}. 1T$^{\prime}$ phase has higher symmetry than 1T$^{\prime\prime}$ phase and satisfies the mirror inversion operation (space group $P_m$). The band structure in Fig.~\hyperref[fig6]{5(d)} shows that its electrons are stronger hopping along the zigzag direction (OB direction) \cite{59} because the band via the Fermi surface undulates significantly in the X-S-$\Gamma$-Y path and is similar to the flat band in the Y-S and $\Gamma$-X paths. 

According to Fig.~\hyperref[fig2]{2(f)}, Re's $d$ orbital mostly contributes to the bands at the Fermi surface, suggesting that it may play an important role in phase transition. In the 1T$^{\prime}$ and 1T$^{\prime\prime}$ phases, Re has octahedral coordination with Se/S, as seen in Fig.~\hyperref[fig6]{5(e)}. Despite variable degrees of distortions, the crystal field effect is still visible. The theory claims that the crystal field causes the 5 $d$ orbitals to split into 3 $t_{2g}$ with lower energy and 2 $e_g$ with higher energy with octahedral coordination \cite{60,61,62}. The outer electrons of Se and S can be written as n$s^2$n$p^4$, whereas Re's are 5$d^5$6$s^2$. Simply put, the electrons in ReSeS rearrange to fill the $p$ orbitals of Se/S, leaving the $d$ orbitals with just 3 electrons, making $t_{2g}$ appear half-filled. When the energy levels of $t_{2g}$ obtain the dispersion relation, the half-filled feature leads the Fermi surface to cross the band, resulting in metallicity. However, this kind of half-filled is not stable because the $t_{2g}$ can be further split into bonding and antibonding orbitals, and the half-filled electrons can fill the bonding orbitals to decrease the energy of the electrons in the system. This process is usually accompanied by structural distortions, meaning an increase in elastic potential energy between ions. When the drop in energy caused by electron recombination is higher than the rise in potential energy, the system will transition to a more stable structure, opening a bandgap. This process is known as the Peierls distortion. By calculating the Helmholtz free energy $\mathcal{F}$ of the system at 0 temperature we find $\mathcal{F}$ of 1T$^\prime$ is about 0.1 eV/atom higher in that of 1T$^{\prime\prime}$ phase, which confirms this process can occur in ReSeS monolayer, and its schematic diagram are displayed in Fig.~\hyperref[fig6]{5(f)}.

\begin{figure*}[!t]
\centering
\includegraphics[width=14cm]{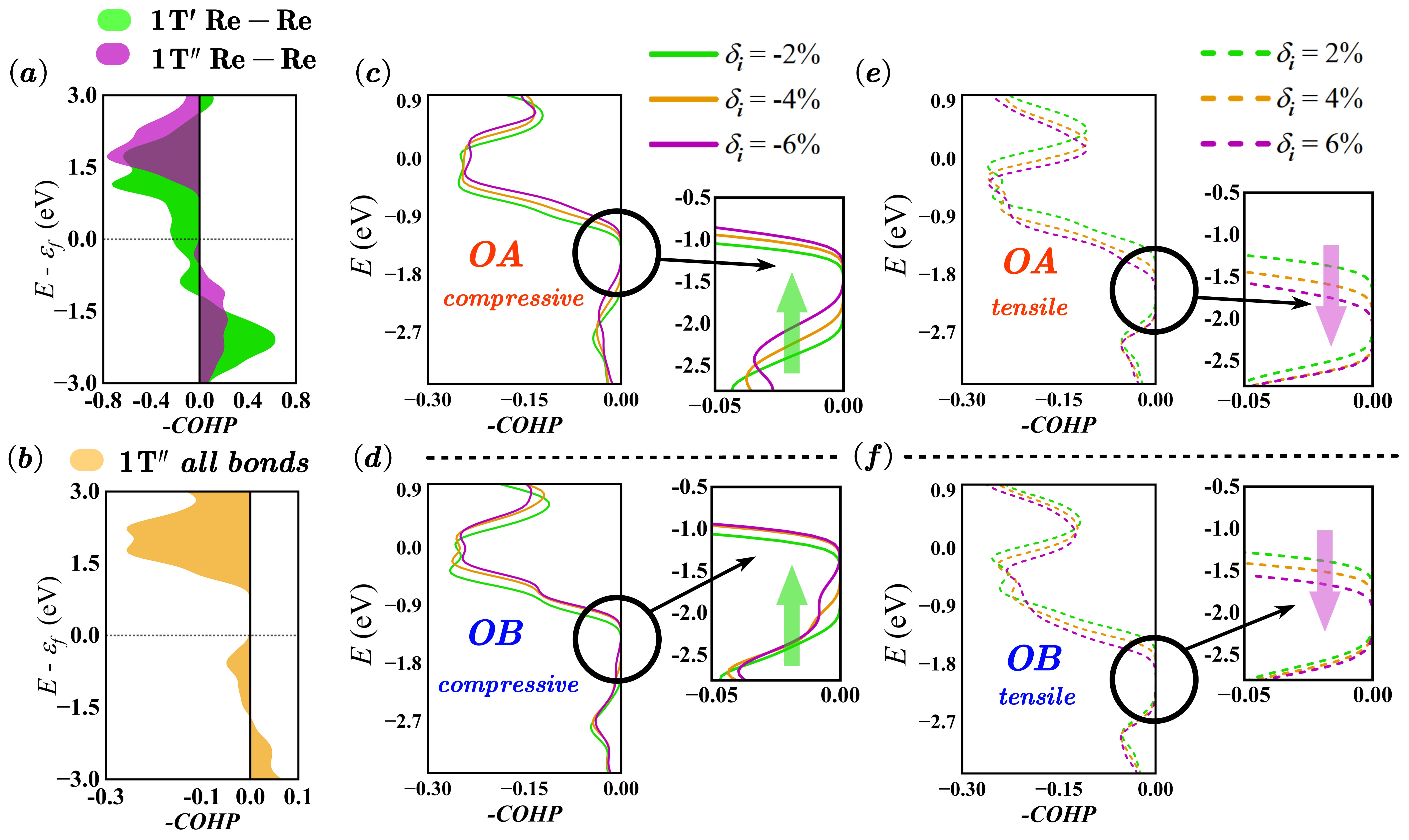}
\caption{\label{fig7}(a) The $-$COHP of Re-Re bonds in 1T$^\prime-$ReSeS (green) and 1T$^{\prime\prime}-$ReSeS (purple) monolayer. (b) the $-$COHP of all bonds in 1T$^{\prime\prime}-$ReSeS monolayer. (c), (d) the $-$COHP of all bonds in 1T$^{\prime\prime}-$ReSeS monolayer with compressive geometric deformations. (e), (f) the $-$COHP of all bonds in 1T$^{\prime\prime}-$ReSeS monolayer with tensile geometric deformations. Black circles and arrows are used to zoom in the region near the Fermi surface. Green, yellow, and purple lines in (c)-(f) represent the degree of geometric deformations. Solid and dot lines in (c)-(f) represent the compressive and tensile deformations, respectively.}
\end{figure*}

To further represent the crystal field effect, we calculate the density of states (DOS) of 1T$^{\prime\prime}-$ReSeS monolayer and the results are in great agreement with our analyses (see Fig.~\hyperref[fig6]{5(f)}). We also ranked the bands corresponding to the 20 Re's $d$ orbitals based on energy, from low to high. Previous analyses indicate that the $d$ orbitals in the 1$st$-12$th$ and 13$th$-20$th$ bands should exhibit $t_{2g}$ and $e_g$ characteristics, respectively, with C/VBM placed in the 6$th$/7$th$ band. orbitals in Fig.~\hyperref[fig6]{5(g), (h)} prove our ideas. $d$ orbitals in the 7$th$ band distribute in the angle between the Re-Se and Re-S bonds and do not face towards the ligand, which is a usual characteristic of $t_{2g}$, whereas $d$ orbitals in the 17$th$ band distribute in the direction of the bonds, having the typical feature of $e_g$. Furthermore, The split of $t_{2g}$ may be observed by the COHP as shown in Fig.~\hyperref[fig7]{6(a)}. A positive/negative value of $-$COHP means bonding/antibonding states. It is clear to find the 1T$^{\prime\prime}$ phase opens a bandgap compared to the 1T$^{\prime}$ due to the split of bonding and antibonding states of Re-Re bonds. All these indicate $d$ orbitals play an important role in Peierls distortion and the creation of the bandgap, and relevant studies also give similar conclusions to corroborate our results \cite{63}.

However, previous studies only take into account the role of Re-Re bonds, and all covalent bonds must be counted to understand the entire bonding situation. Fig.~\hyperref[fig7]{6(b)} show the $-$COHP of all bonds in the ReSeS monolayer. Interestingly, both CBM and VBM exhibit antibonding properties, which shows the presence of antibonding effects between Se/S and Re, as well as between Se and S. This feature remains stable even after applying geometric distortions, as seen in Fig.~\hyperref[fig7]{6(c)-(e)}, and it can also clarify the variations in C/VBM. Because of the poor symmetry, covalent bonds in ReSeS are oriented in no discernible direction, and their distribution in space is isotropic approximately. In other words, it allows us to approximate the fact that geometric deformations applied along any direction compresses or tenses all covalent bonds simultaneously. As a result, when we apply compressive deformations, the system's bond lengths shorten and the antibonding effect becomes greater, increasing energy. Stretching, on the other hand, lengthens bonds, decreasing the effect of antibonding and resulting in a decrease in energy. This then explains the phenomenon in Fig.~\hyperref[fig4]{3(b)} where both C/VBM decrease as the tensile deformations are enhanced, and we also use work function to calculate the value of C/VBM to support these results (see \textit{Supplementary 3}). Similar ideas have also been applied in the analyses of the variation of phosphorene's band structures \cite{64}.

The following question is about the mechanism of the decrease in bandgap. Unlike some TMDs where CBM and VBM change with a relative trend \cite{53}, the CBM and VBM of ReSeS always vary in the same trend. this suggests that the bandgap narrowing is dictated by their relative rates of variation, which can be influenced by the strength of antibonding properties around the CBM and VBM. For the $-$COHP, the strength of antibonding can be visualized in two ways. The first is the integral of $-$COHP, which can be depicted as the area contained by the curve, and the second is the energy, antibonding with higher energy is deemed stronger. As seen in Fig.~\hyperref[fig7]{6(c)-(e)}, antibonding near VBM is always weaker than that near CBM within the range of deformations we applied. Meanwhile, the curve shapes of $-$COHP during geometric deformations change rarely, which shows that in ReSeS, the variations in the strength of antibonding can be evaluated exclusively by the energy. When we apply tensile deformations, the VBM is difficult to decrease further because of its already poor antibonding properties, but the CBM decreases more rapidly, resulting in the narrowing of the bandgap. Compressive deformations, on the contrary, make it harder for the CBM to constantly increase due to its strong antibonding, reducing the bandgap due to the rapid increase of the VBM. Therefore, both compressive and tensile deformations will improve the metallicity.

Actually, this is a kind of saturation-like effect, i.e., the physical quantity cannot increase or decrease indefinitely, but will stabilize around a value or even invert the transformation. Specifically, the CBM of ReSeS with strong antibonding properties will stabilize earlier in compressive deformations, while the VBM with weak antibonding properties will stabilize earlier in tensile deformations, and both are reflected in energy with a slowed rate of change. This can be visualised in the Fig.~\hyperref[fig4]{3(b)} and are also reflected by the results in Fig.~\hyperref[fig7]{6(c)-(e)}. The green arrows in Fig.~\hyperref[fig7]{6(c), (d)} reflects that VBM has a faster rate of change under compressive deformations, while purple arrows in Fig.~\hyperref[fig7]{6(e), (f)} reflects that CBM has a faster rate of change under tensile deformations. compared to the Reduction-Hamiltonian, this understanding provides a more comprehensive and clearer physical image.
\subsection{\label{sec3c}Effective mass and absorption spectrum of electrons}
\begin{figure}[!t]
\centering
\includegraphics[width=14cm]{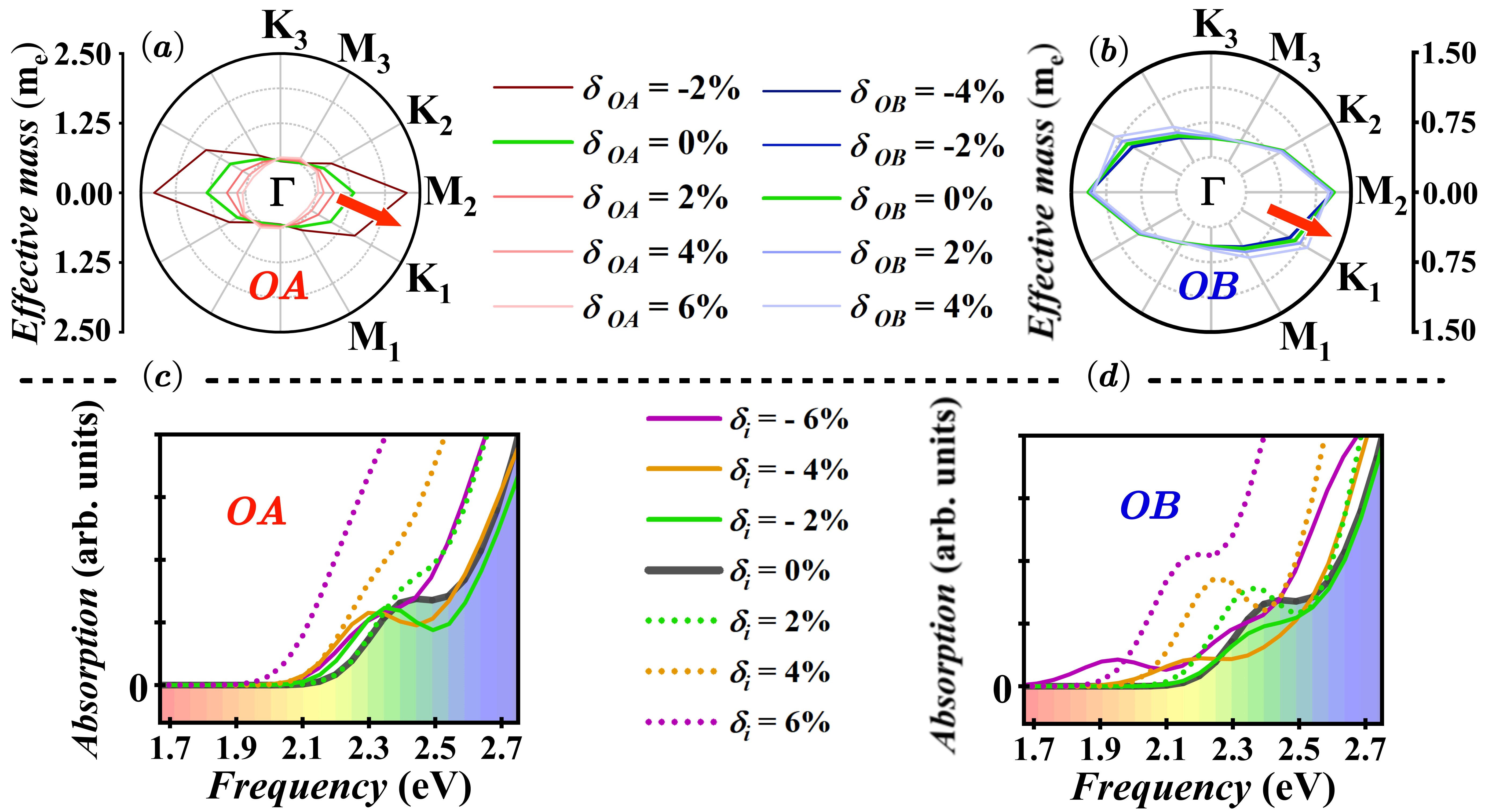}
\caption{\label{fig8}The effective mass of electrons of ReSeS monolayer with geometric deformations along the (a) OA direction and (b) OB direction. The absorption spectrum of electrons of ReSeS monolayer with geometric deformations along the (c) OA direction and (d) OB direction.}
\end{figure}
The discussions above demonstrate that the electronic structures of ReSeS change dramatically when subjected to external distortions. In applications, regulating the deformations allows us to modulate the material's electrical and optical characteristics. The effective mass of electrons \cite{65} plays an important role in electrical transport properties. According to the quasi-classical approximation, the behaviors of electrons in the solid system can be seen as the motion of the wave packet. The dynamical equation of the group velocity of wave packet $\bm{v}$ can be written in Eq.~\ref{eq2}.
\begin{equation}
\dot{\bm{v}}_{\nu}=\sum_{\mu}\bm{F}_{\mu}\left(\mathcal{M}_{\mu\nu}^\ast\right)^{-1}
\label{eq2}
\end{equation}

Where $\mu$, $\nu$ represent the $x$, $y$ and $z$ axis in Cartesian coordinate. $\bm{F}$ means the external force and $\mathcal{M}^\ast$ is the tensor of effective mass, its component can be written as Eq.~\ref{eq3}.
\begin{equation}
\left(\mathcal{M}_{\mu\nu}^\ast\right)^{-1}=\frac{1}{\hbar^2}\partial_{\bm{k}_\mu}\partial_{\bm{k}_\nu}E_{n\bm{k}}
\label{eq3}
\end{equation}

For semiconductors, the effective mass of its conduct band electrons is significant for electrical transport. According to the Eq.~\ref{eq3}, we calculate the effective mass of electrons $m_{\bm{j}}^{\ast}$ at $\Gamma$ point in ReSeS along the directions correspond to high symmetry points ($\textrm{M}_1$, $\textrm{K}_1$, $\textrm{M}_2$, $\textrm{K}_2$, $\textrm{M}_3$, and $\textrm{K}_3$) by $m_{\bm{j}}^{\ast}=\hbar^2/(\partial_{\bm{k_j}}^2E_{n\bm{k}}^{C})_{\bm{k_j}=0}$. $\bm{j}$ means the directions in k-path, and $E_{n\bm{k}}^{C}$ represents the dispersion relation of the conduct band. Results are demonstrated in Fig.~\hyperref[fig8]{7(a), (b)}, and show the great anisotropy. Firstly, $m_{\bm{j}}^{\ast}$ along the $\Gamma-\textrm{M}_2$ and $\Gamma-\textrm{K}_1$ direction (red arrow in Fig.~\hyperref[fig8]{7(a), (b)}) is significantly larger than that in the other directions. In the undeformed situation ($\delta_{\bm{OA}}=\delta_{\bm{OB}}=0\%$), the ratio of maximum $m_{\bm{j}}^{\ast}$ to the minimum is 2.3. And it can further reach to the 4.0 with the $\delta_{\bm{OA}}=-2\%$. Secondly, geometric deformations along the OA direction have a greater impact on the $m_{\bm{j}}^{\ast}$ than distortions along the OB direction, and the $m_{\bm{j}}^{\ast}$ rises dramatically as compressive deformations increase.

Both of these points can be explained by the anisotropic electronic structure of ReSeS. Despite the Peierls distortion, the 1T$^{\prime\prime}$ phase retains the quasi-one-dimensional chain feature in the 1T$^{\prime}$ phase, i.e., the electrons have stronger hopping along the direction of diamond-chain (OB direction), especially the electrons of $d$ orbitals near the Fermi surface. As a result, geometric deformations along the OB direction have a greater effect on the electrical structures. In Fig.~\hyperref[fig4]{3(a)}, for example, distortions along the OB direction cause a more pronounced drop in the bandgap. Similarly, because of the weak wavefunction overlap (nonbonding state) between the diamond-chains along the OA direction, the electrons behave very locally, realizing a larger $m_{\bm{j}}^{\ast}$. Furthermore, compressive deformations along the OA direction will have a greater influence on these nonbonding states, resulting in a large change in the $m_{\bm{j}}^{\ast}$. Especially, compressive deformations strengthens nonbonding states, resulting in a significant increase in $m_{\bm{j}}^{\ast}$ at $\delta_{\bm{OA}}=-2\%$, as previously indicated. Due to deformation potential approximation \cite{52p,66,67,68,69,70,71}, the carrier mobility of 2-D materials along the $\bm{j}$ direction $\tau_{j}$ is positively correlated with $m_{\bm{j}}^{\ast-2}$. Therefore, we can assume that ReSeS has a smaller mobility along the OA direction compared to the OB direction. A similar anisotropy phenomenon is also reported in \cite{21}, indicating the possible use of ReSeS in the design of electrical transport regulable devices.

Meanwhile, the changes in band structures also directly impact the absorption spectrum of electrons. The large forbidden band of ReSeS theoretically truncates its absorption spectrum at energies below the $E_g$, and due to the first principles calculation, the typical peaks in the absorption spectra are more obvious in the region of visible light (from 1.65 eV to 3.10 eV), as shown in Fig.~\hyperref[fig8]{7(c), (d)}. Under the influence of geometric deformations, the narrowing of the bandgap causes a considerable redshift of the spectrum's peaks. Specifically, the deformations along the OA direction make a redshift of the typical absorption peak, changing its energy from 2.35 eV to 2.05 eV during the transition from $\delta_{\bm{OA}}=\pm2\%$ to $\delta_{\bm{OA}}=\pm6\%$. The more significant bandgap narrowing along the OB direction widens this change to the range from 2.40 eV to 1.95 eV, which also reflects the anisotropy in ReSeS monolayer. In the visible light range, this suggests a variation from the blue-green to the orange-yellow region, reflecting the absorption spectrum's strong sensitivity to geometric distortions. Thus, ReSeS's spectroscopic response provides an effective method for monitoring the deformations of this material. 
\section{\label{sec4}Conclusion}

In a nutshell, we report 1T$^{\prime\prime}-$ReSeS monolayer, a kind of TMD with the lowest symmetry in theory (belongs to the points group $P1$). Its double-face structure makes the electrostatic potential exhibit asymmetric features and induces a distinct spectroscopic fingerprint. Following that, we systematically investigate the variations in the band structures under geometric deformations regulation and propose two approaches to understand the mechanism of bandgap decreasing. The Reduction-Hamiltonian greatly simplifies the complex tight-binding Hamiltonian induced by the poor symmetry, but it only provides a qualitative description and does not explain the variation of isolated energy levels $h_i$ with the distortions. The following calculations of COHP claim that the antibonding characteristics of both CBM and VBM cause their energies to decrease with stretching of the system, whereas the difference in strength of antibonding between them leads to a saturation-like effect that causes the VBM to rise faster in compressive deformations and the CBM to fall faster in tensile deformations, which in turn leads to an enhancement of the metallicity of ReSeS under both compressive and tensile distortions. This demonstrates a clearer physical image of the bandgap reduction process. Finally, We investigate the effective mass of electrons and the absorption spectrum with different geometric deformations, both of which exhibit considerable anisotropy. Due to the weak overlap of the $d-$ orbital between the diamond-chains, the dynamics of electrons along the OA direction display strong localization, corresponding to a larger effective mass compared to the OB direction. The narrowing of the bandgap is also reflected in the redshift of the typical peaks of the absorption spectrum, which gives an acceptable way of monitoring the deformations occurring in the material.
\section{Acknowledgments}
Our study is significantly supported by the School of Physics, Nanjing University, and we would like to thank Prof. Haijun Zhang in particular. We acknowledge the Bianshui Riverside Supercomputing Center (BRSC) for computational resources and technical assistance provided by skilled engineers (Wenhan Fang, Kai Hu, and Qiqi Han). We are grateful to Quansheng Wu from the Institute of Physics, Chinese Academy of Science (CAS), for his crucial suggestions. We would also thank Zihao Su from the School of Mathematics, Nanjing University for his help throughout our hard times.
\section{CRediT authorship contribution statement}
$\textbf{Timsy Tinche Lin}$: Writing – original draft, Software, Investigation, Methodology, Formal analysis, Validation. $\textbf{Haochen Deng}$: Writing – review and editing, Software, Resources, Data curation. $\textbf{Junwei Ma}$: Visualization, Conceptualization, Data curation. $\textbf{Lizhe Liu}$: Writing – review and editing, Supervision.

\bibliographystyle{elsarticle-num} 
\bibliography{elsarticle-template-num}






\end{document}